\def\draft{0}  
\documentclass[proceedings]{stacs}
\stacsheading{2010}{119-130}{Nancy, France}
 \firstpageno{119}

\usepackage{latexsym}
\usepackage{amssymb}
\usepackage{times}
\usepackage{amsmath,amsfonts,amsthm}
\usepackage{enumerate}
\usepackage{clrscode}
\usepackage{url}

\theoremstyle{plain}
\theoremstyle{definition}

\ifnum\draft=1
\newcommand{\Lnote}[1]{{[\bf Liu's Note: #1]}}
\newcommand{\Cnote}[1]{{[\bf Chung's Note: #1]}}
\newcommand{\Bnote}[1]{{[\bf Braverman's Note: #1]}}
\else
\newcommand{\Lnote}[1]{{}}
\newcommand{\Cnote}[1]{{}}
\newcommand{\Bnote}[1]{{}}
\fi

\newcommand{\vecv}{{\vec{v}} }
\newcommand{\barS}{{\bar{S}} }
\newcommand{\bfp}{{\mathbf p} }
\newcommand{\bfq}{{\mathbf q} }
\newcommand{\bfa}{{\mathbf a} }
\newcommand{\bfb}{{\mathbf b} }
\newcommand{\bfc}{{\mathbf c} }
\newcommand{\bfd}{{\mathbf d} }
\newcommand{\Var}{{\mathrm{Var}} }
\newcommand{\E}{{\mathrm{E}} }

\begin{document}

\title[AMS Without $\bf{4}$-Wise Independence on Product Domains]{AMS Without $\bf{4}$-Wise Independence on Product Domains\footnote{This paper is a merge from the  work of \cite{BO01, BO02, CLM}}}

\author[lab1]{V. Braverman}{Vladimir Braverman}
\address[lab1]{University of California Los Angeles. Supported in part by NSF grants 0716835, 0716389, 0830803, 0916574
and Lockheed Martin Corporation.}  
\email{vova@cs.ucla.edu}  
\urladdr{\url{http://www.cs.ucla.edu/~vova}}  

\author[lab2]{K. Chung}{Kai-Min Chung}
\address[lab2]{Harvard School of Engineering and Applied Sciences. Supported by US-Israel BSF grant 2006060 and NSF grant CNS-0831289.}	
\email{kmchung@fas.harvard.edu}  
\urladdr{\url{http://people.seas.harvard.edu/~kmchung/}}  

\author[lab3]{Z. Liu}{Zhenming Liu}
\address[lab3]{Harvard School of Engineering and Applied Sciences. Supported in part by NSF grant CNS-0721491. The work was finished during an internship in Microsoft Research Asia.}
\email{zliu@fas.harvard.edu}  
\urladdr{\url{http://people.seas.harvard.edu/~zliu/}}  

\author[lab4]{M. Mitzenmacher}{Michael Mitzenmacher}
\address[lab4]{Harvard School of Engineering and Applied Sciences. Supported in part by NSF grant CNS-0721491 and research grants from Yahoo!, Google, and Cisco.}
\email{michaelm@eecs.harvard.edu}  
\urladdr{\url{http://www.eecs.harvard.edu/~michaelm/}}  

\author[lab5]{R. Ostrovsky}{Rafail Ostrovsky}
\address[lab5]{University of California Los Angeles. Supported in part by IBM Faculty Award, Lockheed-Martin Corporation
Research Award, Xerox Innovation Group Award, the Okawa Foundation Award,
Intel, Teradata, NSF grants 0716835, 0716389, 0830803, 0916574 and U.C.
MICRO grant.}  
\email{rafail@cs.ucla.edu}  
\urladdr{\url{http://www.cs.ucla.edu/~rafail}}  




\begin{abstract}
In their seminal work, Alon, Matias, and Szegedy introduced
several sketching techniques, including showing that $4$-wise
independence is sufficient to obtain good approximations of the second
frequency moment.  In this work, we show that their sketching
technique can be extended to product domains $[n]^k$ by using the
product of $4$-wise independent functions on $[n]$.
Our work extends that of Indyk and McGregor, who showed the result
for $k = 2$.  Their primary motivation was the problem of identifying correlations in data streams.
In their model, a stream of pairs $(i,j) \in [n]^2$ arrive,
giving a joint distribution $(X,Y)$, and they find approximation
algorithms for how close the joint distribution is to the product of
the marginal distributions under various metrics, which naturally
corresponds to how close $X$ and $Y$ are to being independent.
By using our technique, we obtain a new result for the problem of
approximating the $\ell_2$ distance between the
joint distribution and the product of the marginal distributions for $k$-ary
vectors, instead of just pairs, in a single pass.
Our analysis gives a randomized algorithm that is a $(1
\pm \epsilon)$ approximation (with probability $1-\delta$) that
requires space logarithmic in $n$ and $m$ and proportional to $3^k$.
\vspace{-0.6cm}
\end{abstract}

\maketitle

\section{Introduction}

In their seminal work, Alon, Matias and Szegedy \cite{ams} presented
celebrated sketching techniques and showed that $4$-wise independence
is sufficient to obtain good approximations of the second frequency
moment.  Indyk and McGregor \cite{IM08} make use of this technique in
their work introduce the problem of measuring independence in the
streaming model.  There they give efficient algorithms for
approximating pairwise independence for the $\ell_1$ and $\ell_2$ norms.
In their model, a stream of pairs $(i,j) \in [n]^2$ arrive, giving a
joint distribution $(X,Y)$, and the notion of approximating pairwise
independence corresponds to approximating the distance between the
joint distribution and the product of the marginal distributions for
the pairs.  Indyk and McGregor state, as an explicit open question in
their paper, the problem of whether one can estimate $k$-wise
independence on $k$-tuples for any $k> 2$.  In particular, Indyk and McGregor show that, for the $\ell_2$ norm, they
can make use of the product of $4$-wise independent functions on $[n]$
in the sketching method of Alon, Matias, and Szegedy.  We extend their
approach to show that on the product domain $[n]^k$, the sketching
method of Alon, Matias, and Szegedy works when using the product of
$k$ copies of $4$-wise independent functions on $[n]$.  The cost is that the
memory requirements of our approach grow exponentially with $k$,
proportionally to $3^k$.


Measuring
independence and $k$-wise independence is a fundamental problem with
many applications (see e.g., Lehmann \cite{stat}). Recently, this problem was also addressed in other models by, among
others, Alon, Andoni, Kaufman, Matulef, Rubinfeld and Xie
\cite{k-wise_independenc}; Batu, Fortnow, Fischer, Kumar, Rubinfeld
and White \cite{batu_independence}; Goldreich and Ron \cite{ind1};
Batu, Kumar and Rubinfeld \cite{ind2}; Alon, Goldreich and Mansour
\cite{ind3}; and Rubinfeld and Servedio \cite{ind4}.
Traditional non-parametric methods of testing independence over empirical data
usually require space complexity that is polynomial to either the support
size or input size. The scale of contemporary data sets often
prohibits such space complexity.  It is therefore natural to ask
whether we will be able to design algorithms to test for independence
in streaming model. Interestingly, this specific problem appears not
to have been introduced until the work of Indyk and McGregor.  While
arguably results for the $\ell_1$ norm would be stronger than for
the $\ell_2$ norm in this setting, the problem for $\ell_2$ norms
is interesting in its own right. The problem for the $\ell_1$ norm has been
recently resolved by Braverman and Ostrovsky in \cite{BO03}. They gave an $(1\pm \epsilon, \delta)$-approximation algorithm
that makes a single pass over a data stream and uses polylogarithmic memory.

\subsection{Our Results}
In this paper we generalize the ``sketching of sketches''
result of Indyk and McGregor.
Our specific theoretical contributions can be summarized as follows:

\ \

\noindent
\emph{\textbf{Main Theorem. }}

\noindent
Let $\vecv \in \mathrm R^{(n^k)}$ be a vector with entries $\vecv_{\bfp} \in R$ for $\bfp \in [n]^k$.
Let $h_1,\dots,h_k:[n] \rightarrow \{-1, 1\}$ be independent copies of 4-wise independent hash functions;  that is, $h_i(1),\dots,h_i(n) \in \{-1,1\}$ are $4$-wise independent hash functions for each $i\in[k]$, and $h_1(\cdot),\dots,h_k(\cdot)$ are mutually independent.  Define $H(\bfp) = \prod_{i =1}^k h_j(p_j)$, and the sketch $Y = \sum_{p\in [n]^k} \vec v_{\mathbf p} H(p)$.

We prove that the sketch $Y$ can be used to give an efficient approximation for $\|\vecv\|^2$; our result is stated formally in Theorem \ref{tm:main1}. Note that $H$ is not $4$-wise independent.

\ \ \


As a corollary, the main application of our main theorem is to extend the result of Indyk and McGregor \cite{IM08} to detect the dependency of $k$ random variables in streaming model.

\begin{corollary}
For every $\epsilon > 0$ and $\delta > 0$, there exists a randomized algorithm that computes,
given a sequence $a_1, \dots, a_m$ of $k$-tuples, in one pass and using
$O(3^k\epsilon^{-2}\log \frac 1 \delta(\log m + \log n))$ memory bits,
a number $Y$ so that the probability $Y$ deviates from the $\ell_2$
distance between product and joint distribution by more than a factor of
$(1+\epsilon)$ is at most $\delta$.
\end{corollary}

\subsection{Techniques and a Historical Remark}

This paper is merge from \cite{BO01, BO02, CLM}, where the same result was obtained with different
proofs. The proof of \cite{CLM} generalizes the geometric approach of
Indyk and McGregor \cite{IM08} with new geometric observations.  The
proofs of \cite{BO01,BO02} are more combinatorial in nature. These papers
offer new insights, but due to the space limitation, we focus
on the proof from \cite{BO02} in this paper. Original papers are
available on line and are recommended to the interested reader.

\section{The Model}
We provide the general underlying model.  Here we mostly follow the notation of \cite{BO01,IM08}.

Let $S$ be a stream of size $m$ with elements $a_1, \dots, a_m$, where
$a_i \equiv (a^1_i, \dots, a^k_i) \in [n]^k$.
(When we have a sequence of elements that are themselves vectors, we
denote the sequence number by a subscript and the vector entry by a
superscript when both are needed.)
The stream $S$ defines an
\emph{empirical} distribution over $[n]^k$ as follows: the frequency
$f(\omega)$ of an element $\omega \in [n]^k$ is defined as the number
of times it appears in $S$, and the empirical distribution is
$$\Pr[\omega] = \frac{f(\omega)}{m} \quad \mbox{for any $\omega \in [n]^k$.}$$

Since $\omega = (\omega_1, \dots, \omega_k)$ is a vector of size $k$, we
may also view the streaming data as defining a joint distribution over
the random variables $X_1, \dots, X_k$ corresponding to the values in
each dimension. (In the case of $k = 2$, we write the random variables
as $X$ and $Y$ rather than $X_1$ and $X_2$.) There is a natural way of
defining marginal distribution for the random variable $X_i$: for
$\omega_i \in [n]$, let $f_i(\omega_i)$ be the number of times
$\omega_i$ appears in the $i$th coordinate of an element of $S$,
or
$$f_i(\omega_i) = \left|\{a_j \in S: a^i_j = \omega_i\}\right|.$$
The empirical marginal distribution $\Pr_i[\cdot]$  for the $i$th coordinate is defined as
$$\mathrm{Pr}_i[\omega_i] = \frac{f_i(\omega_i)}{m} \quad \mbox{for any $\omega_i \in [n]$.}$$

Next let $\vec v$ be the vector in ${\mathrm R}^{[n]^k}$ with $\vec v_\omega = \Pr[\omega] - \prod_{1 \leq i\leq k}\Pr_i[\omega_i]$ for all $\omega \in [n]^k$. Our goal is to approximate the value
\begin{equation}\label{eqn:alpha}
\|\vec v\| \equiv \left(\sum_{\omega \in [n]^k}\left|\Pr[\omega] - \prod_{1 \leq i \leq k}\mathrm{Pr}_i[\omega_i] \right|^2\right)^{\frac 1 2}.
\end{equation}
This represent the $\ell_2$ norm between the tensor of the marginal distributions and the joint distribution,
which we would expect to be close to zero in the case where the $X_i$ were truly independent.

Finally, our algorithms will assume the availability of 4-wise independent hash functions.
For more on 4-wise independence, including efficient implementations, see \cite{BCH, TZ04}.
For the purposes of this paper, the following simple definition will suffice.

\begin{definition}\emph{(4-wise independence)}
A family of hash functions $\mathcal H$ with domain $[n]$ and range $\{-1, 1\}$ is \emph{4-wise independent} if for any distinct values
$i_1, i_2, i_3, i_4 \in [n]$ and any $b_1, b_2, b_3, b_4 \in \{-1, 1\}$, the following equality holds,
$$\Pr_{h \leftarrow \mathcal H}\left[h(i_1) = b_1, h(i_2) = b_2, h(i_3) = b_3, h(i_4) = b_4\right] = 1/16.$$
\end{definition}
\begin{remark}In \cite{IM08}, the family of 4-wise independent hash functions $\mathcal H$ is called 4-wise independent random vectors. For consistencies within our paper, we will always view the object $\mathcal H$ as a hash function family.
\end{remark}

\section{The Algorithm and its Analysis for $k=2$} \label{sec:kis2}
We begin by reviewing the approximation algorithm and associated proof
for the $\ell_2$ norm given in \cite{IM08}. Reviewing this result
will allow us to provide the necessary notation and frame the setting
for our extension to general $k$.  Moreover, in our proof, we find
that a constant in Lemma 3.1 from \cite{IM08} that we subsequently
generalize appears incorrect.  (Because of this, our proof is slightly
different and more detailed than the original.)  Although the error is
minor in the context of their paper (it only affects the constant
factor in the order notation), it becomes more important when
considering the proper generalization to larger $k$, and hence it
is useful to correct here.

In the case $k=2$, we assume that the sequence $(a^1_1, a^2_1), (a^1_2, a^2_2), \dots, (a^1_m, a^2_m)$ arrives an item by an item.
Each $(a^1_i, a^2_i)$ (for $1 \leq i \leq m$) is an element in $[n]^2$. The random variables $X$ and $Y$ over $[n]$ can be expressed
as follows:
\begin{displaymath}
\left\{
\begin{array}{lllll}
\Pr[i, j] & = & \Pr[X = i, Y = j] & = & |\{\ell: (a^1_\ell, a^2_\ell) = (i, j)\}| / m \\
\Pr_1[i] & = & \Pr[X = i] & = & |\{\ell: (a^1_\ell, a^2_\ell) = (i, \cdot)\}| / m \\
\Pr_2[j] & = & \Pr[Y = j] & = & |\{\ell: (a^1_\ell, a^2_\ell) = (\cdot, j)\}| / m.
\end{array}\right.
\end{displaymath}
We simplify the notation and use
$p_i \equiv \Pr[X = i]$, $q_j \equiv \Pr[Y = j]$,
$r_{i,j} = \Pr[X = i, Y = j]$.
and
$s_{i,j} = \Pr[X = i]\Pr[Y = j]$.

Indyk and McGregor's algorithm proceeds in a similar fashion
to the streaming algorithm presented in \cite{ams}. Specifically let $s_1 = 72\epsilon^{-2}$ and $s_2 = 2\log (1/\delta)$.
The algorithm computes $s_2$ random variables $Y_1, Y_2, \dots, Y_{s_2}$ and
outputs their median. The output is the algorithm's estimate on the norm of $v$ defined in Equation~\ref{eqn:alpha}.
Each $Y_i$ is the average of $s_1$ random variables $Y_{ij}$: $1 \leq j \leq s_1$, where $Y_{ij}$ are independent, identically
distributed random variables. Each of the variables $D = D_{ij}$ can be computed from the algorithmic routine shown in
Figure~\ref{fig:algx}.

\begin{figure}[htp]
\begin{codebox}
\Procname{$\proc{2-D Approximation}\left((a^1_1, a^2_1), \dots, (a^1_m, a^2_m)\right)$}
\li Independently generate 4-wise independent random functions $h_1, h_2$ from $[n]$ to $\{-1, 1\}$.
\li \For $c \gets 1$ \To $m$
\li \Do Let the $c$th item $(a^1_c, a^2_c) = (i, j)$
\li $t_1 \gets t_1 + h_1(i)h_2(j)$, $t_2 \gets t_2 + h_1(i)$, $t_3 \gets t_3 + h_2(j)$.
\End
\li Return $Y = (t_1 / m - t_2t_3 / m^2)^2$.
\end{codebox}
\caption{ The procedure for generating random variable $Y$ for $k=2$.}
\label{fig:algx}
\end{figure}

\noindent
By the end of the process $\proc{2-D Approximation}$, we have 
$t_1 / m = \sum_{i, j \in [n]}h_1(i)h_2(j)r_{i,j}$,
$t_2 / m = \sum_{i \in [n]}h_1(i)p_i$, and $t_3 / m = \sum_{i \in [n]}h_2(i)q_i$. Also, when a vector is in ${\mathrm R}^{(n^2)}$, its indices can be represented by $(i_1, i_2) \in [n]^2$. In what follows, we will use a bold letter to represent the index of a high dimensional vector, e.g., $v_{\mathbf i} \equiv v_{i_1, i_2}$.
The following Lemma shows that the expectation of $Y$ is $\|v\|^2$ and the variance of $Y$ is at most $8(\mathrm E[Y])^2$ because $\mathrm E[Y^2] \leq 9 \mathrm E[Y]^2$.

\begin{lemma}\label{lem:varx}(\cite{IM08}) Let $h_1, h_2$ be two independent instances of 4-wise independent hash functions from $[n]$ to $\{-1, 1\}$. Let $v \in \mathrm R^{n^2}$ and $H(\mathbf i)(\equiv H\big((i_1, i_2)\big) = h_1(i_i)\cdot h_2(i_2)$. Let us define $Y = \left(\sum_{\mathbf i \in [n]^2}H(\mathbf i)v_{\mathbf i}\right)^2$. Then
$\mathrm E[Y] = \sum_{\mathbf i \in [n]^2}\vec v_{\mathbf i}^2$ and $\mathrm E[Y^2] \leq 9(\mathrm E[Y])^2$, which implies $\mathrm{Var}[Y] \leq 8 E^2[Y]$.
\end{lemma}

\begin{proof}
We have $\mathrm E[Y] = \mathrm E[(\sum_{\mathbf i}H(\mathbf
i)\vec v_{\mathbf i})^2] = \sum_{\mathbf i}\vec v^2_{\mathbf i} \mathrm
E[H^2(\mathbf i)] + \sum_{\mathbf i \neq \mathbf j}\vec v_{\mathbf
i}\vec v_{\mathbf j}\mathrm E[H(\mathbf i)H(\mathbf j)]$.  For all
$\mathbf i \in [n]^2$, we know $h^2(\mathbf i) = 1$. On the other
hand, $H(\mathbf i)H(\mathbf j) \in \{-1, 1\}$.  The probability
that $H(\mathbf i)H(\mathbf j) = 1$ is $\Pr[H(\mathbf i)H(\mathbf
j) = 1] = \Pr[h_1(i_1)h_1(j_1)h_2(i_2)h_2(j_2) = 1] = 1/16 +
\binom{4}{2}1/16 + 1/16 = 1/2$. The last equality holds is because
$h_1(i_1)h_1(j_1)h_2(i_2)h_2(j_2) = 1$ is equivalent to saying either all
these variables are 1, or exactly two of these variables are -1, or all
these variables are -1. Therefore, $\mathrm E[h(\mathbf i)h(\mathbf
j)] = 0$. Consequently, $\mathrm E[Y] = \sum_{\mathrm i \in
[n]^2}(\vec v_{\mathrm i})^2$.

Now we bound the variance. Recall that $\mathrm{Var}[Y] = \mathrm E[Y^2] - \mathrm E[Y]^2$, we bound
$$\mathrm E[Y^2] = \sum_{\mathbf i, \mathbf j, \mathbf k, \mathbf l \in [n]^2}
\mathrm E[H(\mathbf i)H(\mathbf j)H(\mathbf k)h(\mathbf l)]\vec v_{\mathbf i}\vec v_{\mathbf j}\vec v_{\mathbf k}\vec v_{\mathbf l} \leq
 \sum_{\mathbf i, \mathbf j, \mathbf k, \mathbf l \in [n]^2} \left|\mathrm E[H(\mathbf i)H(\mathbf j)H(\mathbf k)H(\mathbf l)]\right| \cdot |\vec v_{\mathbf i}\vec v_{\mathbf j}\vec v_{\mathbf k}\vec v_{\mathbf l}|.$$

Also $\left|\mathrm E[H(\mathbf i)H(\mathbf j)H(\mathbf k)H(\mathbf l)]\right| \in \{0, 1\}$. The quantity
$\mathrm E[H(\mathbf i)H(\mathbf j)H(\mathbf k)H(\mathbf l)] \neq 0$ if and only if the following relation holds,
\begin{equation}\label{eqn:relation}
\forall s \in [2]: \left( (i_s = j_s) \wedge (k_s = l_s) \right) \vee \left( (i_s = k_s) \wedge (j_s = l_s) \right) \vee \left( (i_s = l_s) \wedge (k_s = j_s) \right).
\end{equation}
Denote the set of 4-tuples $(\mathbf i, \mathbf j, \mathbf k, \mathbf l)$ that satisfy the above relation by $\mathcal D$.
We may also view each 4-tuple as an ordered set that consists of 4 points in $[n]^2$.
Consider the unique smallest axes-parallel rectangle in $[n]^2$
that contains a given 4-tuple in $\mathcal D$ (i.e. contains the
four ordered points).  Note this could either be a (degenerate) line segment
or a (non-degenerate) rectangle, as we discuss below.
Let $M: \mathcal D \rightarrow \{A, B, C, D\}$ be the function
that maps an element $\sigma \in \mathcal D$ to the smallest rectangle
$ABCD$ defined by $\sigma$. Since a
rectangle can be uniquely determined by its diagonals, we may write
$M: \mathcal D \rightarrow (\chi_1, \chi_2, \varphi_1, \varphi_2)$,
where $\chi_1 \leq \chi_2 \in [n]$, $\varphi_1 \leq \varphi_2 \in [n]$
and the corresponding rectangle is understood to be the one with
diagonal $\{(\chi_1, \varphi_1), (\chi_2, \varphi_2)\}$. Also, the
inverse function $M^{-1}(\chi_1, \chi_2, \varphi_1, \varphi_2)$
represents the pre-images of $(\chi_1, \chi_2, \varphi_1, \varphi_2)$
in $\mathcal D$. $(\chi_1, \chi_2, \varphi_1, \varphi_2)$ is degenerate if either $\chi_1 =
\chi_2$ or $\varphi_1 = \varphi_2$, in which case the rectangle (and
its diagonals) correspond to the segment itself, or $\chi_1 =
\chi_2$ and $\varphi_1 = \varphi_2$, and the rectangle is just a
single point.

\begin{example}
Let $\mathbf i = (1, 2)$, $\mathbf j = (3, 2)$,
$\mathbf k = (1, 5)$, and $\mathbf l = (3, 5)$. The tuple is in $\mathcal D$ and its corresponding bounding rectangle is
a non-degenerate rectangle. The function $M(\mathbf i, \mathbf j, \mathbf k, \mathbf l) = (1, 3, 2, 5)$.
\end{example}

\begin{example} Let $\mathbf i = \mathbf j = (1, 4)$
and $\mathbf k = \mathbf l = (3, 7)$. The tuple is also in $\mathcal D$ and
minimal bounding rectangle formed by these points is an interval $\{(1, 4), (3, 7)\}$. The function
$M(\mathbf i, \mathbf j, \mathbf k, \mathbf l) = (1, 3, 4, 7)$.
\end{example}

To start we consider the non-degenerate cases.
Fix any $(\chi_1, \chi_2, \varphi_1, \varphi_2)$ with $\chi_1 < \chi_2$ and $\phi_1 < \phi_2$.
There are in total $\binom{4}{2}^2 = 36$ tuples $(\mathbf i,
\mathbf j, \mathbf k, \mathbf l)$ in $\mathcal D$ with $M(\mathbf i,
\mathbf j, \mathbf k, \mathbf l) = (\chi_1, \chi_2, \varphi_1, \varphi_2)$.  Twenty-four of these tuples
correspond to the setting where none of $\mathbf i, \mathbf j, \mathbf k, \mathbf l$ are equal, as there
are twenty-four permutations of the assignment of the labels $\mathbf i, \mathbf j, \mathbf k, \mathbf l$ to the four points. (This corresponds to the first example).  In this
case the four points form a rectangle, and we have
$|\vec v_{\mathbf i}\vec v_{\mathbf j}\vec v_{\mathbf k}\vec v_{\mathbf l}| \leq \frac 1 2((\vec v_{\chi_1, \varphi_1}\vec v_{\chi_2, \varphi_2})^2 + (
\vec v_{\chi_1, \varphi_2} \vec v_{\chi_2, \varphi_1})^2)$.
Intuitively, in these cases, we assign the ``weight'' of the tuple to the diagonals.

The remaining twelve tuples in $M^{-1}(\chi_1, \chi_2, \varphi_1, \varphi_2)$ correspond to intervals. (This corresponds to the second example.)
In this case two of $\mathbf i, \mathbf j, \mathbf k, \mathbf l$ correspond to one endpoint of the interval, and the other two labels
correspond to the other endpoint.
Hence we have either $|\vec v_{\mathbf i}\vec v_{\mathbf j}\vec v_{\mathbf k}\vec v_{\mathbf l}|
= (\vec v_{\chi_1, \varphi_1}\vec v_{\chi_2, \varphi_2})^2$ or $|\vec v_{\mathbf i}\vec v_{\mathbf j}\vec v_{\mathbf k}\vec v_{\mathbf l}| = (\vec v_{\chi_1, \varphi_2}\vec v_{\chi_2, \varphi_1})^2$, and there are six tuples for each case.

Therefore for any $\chi_1 < \chi_2 \in [n]$
and $\varphi_1 < \varphi_2 \in [n]$ we have:
$$\sum_{\scriptsize{}\substack{(\mathbf i, \mathbf j, \mathbf k, \mathbf l) \in \\  M^{-1}(\chi_1, \chi_2,\varphi_1, \varphi_2)}}\hspace{-.5cm} |v_{\mathbf i}v_{\mathbf j}v_{\mathbf k}v_{\mathbf l}|
\leq 18((v_{\chi_1, \varphi_1}v_{\chi_2, \varphi_2})^2 + (
v_{\chi_1, \varphi_2}, v_{\chi_2, \varphi_1})^2).$$

The analysis is similar for the degenerate cases, where the constant 18 in the bound above is now quite loose.
When exactly one of
$\chi_1 = \chi_2$ or $\varphi_1 = \varphi_2$ holds, the size of
$M^{-1}(\chi_1, \chi_2, \varphi_1, \varphi_2)$ is $\binom{4}{2} =
6$, and the resulting intervals correspond to vertical or horizontal
lines.  When both $\chi_1 = \chi_2$ and $\varphi_1 = \varphi_2$, then
$|M^{-1}(\chi_1,\chi_2, \varphi_1, \varphi_2)| = 1$. In sum, we have
Following the same analysis as for the non-degenerate cases, we find

$$
\sum_{\mathbf i, \mathbf j, \mathbf k, \mathbf l \in \mathcal D} |\vec v_{\mathbf i}\vec v_{\mathbf j}\vec v_{\mathbf k}\vec v_{\mathbf l}|
 =  \sum_{\scriptsize{}\substack{\chi_1 \leq \chi_2 \\ \varphi_1 \leq \varphi_2}}\sum_{\scriptsize{}\substack{(\mathbf i, \mathbf j, \mathbf k, \mathbf l) \in \\  M^{-1}(\chi_1,\chi_2,\varphi_1, \varphi_2)}}\hspace{-.5cm} |\vec v_{\mathbf i}\vec v_{\mathbf j}\vec v_{\mathbf k}\vec v_{\mathbf l}| \\
$$
$$
\leq \sum_{\scriptsize{}\substack{\chi_1 < \chi_2 \\ \varphi_1 < \varphi_2}} 18((\vec v_{\chi_1, \varphi_1}\vec v_{\chi_2, \varphi_2})^2 + (
\vec v_{\chi_1, \varphi_2}\vec v_{\chi_2, \varphi_1})^2) + \sum_{\scriptsize{}\substack{\chi_1 = \chi_2 \\ \varphi_1 < \varphi_2}} 6((\vec v_{\chi_1, \varphi_1}\vec v_{\chi_2, \varphi_2})^2 + (
\vec v_{\chi_1, \varphi_2}\vec v_{\chi_2, \varphi_1})^2) 
$$
$$
+ \sum_{\scriptsize{}\substack{\chi_1 < \chi_2 \\ \varphi_1 = \varphi_2}} 6((\vec v_{\chi_1, \varphi_1}\vec v_{\chi_2, \varphi_2})^2 + (
\vec v_{\chi_1, \varphi_2} \vec v_{\chi_2, \varphi_1})^2) + \sum_{\scriptsize{}\substack{\chi_1 = \chi_2 \\ \varphi_1 = \varphi_2}} (\vec v_{\chi_1, \varphi_1}\vec v_{\chi_2, \varphi_2})^2 \\
$$
$$
\leq 9 \sum_{\scriptsize{}\substack{\mathbf i \in [n]^2 \\ \mathbf j \in [n]^2}}(\vec v_{\mathbf i}\vec v_{\mathbf j})^2 = 9 \mathrm E^2[Y].
$$

Finally, we have $\sum_{\mathbf i, \mathbf j, \mathbf k, \mathbf l \in [n]^2} \left|\mathrm E[H(\mathbf i)H(\mathbf j)H(\mathbf k)H(\mathbf l)]\right| \cdot |\vec v_{\mathbf i}\vec v_{\mathbf j}\vec v_{\mathbf k}\vec v_{\mathbf l}|
\leq \sum_{\mathbf i, \mathbf j, \mathbf k, \mathbf l \in \mathcal D} |\vec v_{\mathbf i}\vec v_{\mathbf j}\vec v_{\mathbf k}\vec v_{\mathbf l}| \leq 9 \mathrm E^2[Y]$ and
$\mathrm{Var}[Y] \leq 8\mathrm E[Y]^2$.
\end{proof}

We emphasize the geometric interpretation of the above proof as follows. The goal is to bound the variance by a constant times $\mathrm E^2[Y] = \sum_{\scriptsize{}\substack{\mathbf i,\mathbf j \in [n]^2 }}(\vec v_{\mathbf i}v_{\mathbf j})^2$, where the index set is the set of all possible lines in plane $[n]^2$ (each line appears twice). We first show that $\mathrm{Var}[Y] \leq \sum_{\mathbf i, \mathbf j, \mathbf k, \mathbf l \in \mathcal D} |\vec v_{\mathbf i}\vec v_{\mathbf j}\vec v_{\mathbf k}\vec v_{\mathbf l}|$,
where the 4-tuple index set corresponds to a set of rectangles in a natural way. The main idea of \cite{IM08} is to use inequalities of the form $|\vec v_{\mathbf i}\vec v_{\mathbf j}\vec v_{\mathbf k}\vec v_{\mathbf l}| \leq \frac 1 2((\vec v_{\chi_1, \varphi_1}\vec v_{\chi_2, \varphi_2})^2 + (\vec v_{\chi_1, \varphi_2}\vec v_{\chi_2, \varphi_1})^2)$ to assign the ``weight'' of each $4$-tuple to the diagonals of the corresponding rectangle.  The above analysis shows that $18$ copies of all lines are sufficient to accommodate all 4-tuples. While similar inequalities could also assign the weight of a $4$-tuple to the vertical or horizontal edges of the corresponding rectangle, using vertical or horizontal edges is problematic. The reason is that there are $\Omega(n^4)$ $4$-tuples but only $O(n^3)$ vertical or horizontal edges, so some lines would receive $\Omega(n)$ weight, requiring $\Omega(n)$ copies. This problem is already noted in \cite{BO01}.


Our bound here is $\mathrm E[Y^2] \leq 9\mathrm E^2[Y]$, while in
\cite{IM08} the bound obtained is $\mathrm E[Y^2] \leq 3\mathrm
E^2[Y]$. There appears to have been an error in the derivation in \cite{IM08};
some intuition comes from the following example.
We note that $|\mathcal D|$ is at
least $\binom{4}{2}^2 \cdot \binom{n}{2}^2 = 9n^4 - 9n^2$. (This counts the number of non-degenerate $4$-tuples.) Now if we
set $v_i = 1$ for all $1 \leq i \leq n^2$, we have $\mathrm E[Y^2]
\geq |\mathcal D| = 9n^4 - 9n^2 \sim \mathrm 9 \mathrm E^2(D)$, which
suggests $\mathrm{Var}[D] > 3 \mathrm E^2[D]$. Again, we emphasize this
discrepancy is of little importance to \cite{IM08};  the point there is that
the variance is bounded by a constant factor times the square of the expectation.
It is here, where we are generalizing to $k \geq 3$, that the exact
constant factor is of some importance.

Given the bounds on the expectation and variance for the $D_{i,j}$,
standard techniques yield a bound on the performance
of our algorithm.

\begin{theorem}\label{thm:case2} For every $\epsilon > 0$ and $\delta > 0$, there exists a randomized algorithm
that computes, given a sequence $(a^1_1, a^2_1), \dots, (a^1_m, a^2_m)$, in one pass and using $O(\epsilon^{-2}\log \frac 1 \delta(\log m + \log n))$
memory bits, a number $\mathrm{Med}$ so that the probability $\mathrm{Med}$ deviates from $\|v\|^2$ by more than $\epsilon$ is at most $\delta$.
\end{theorem}
\begin{proof}
Recall the algorithm described in the beginning of Section \ref{sec:kis2}: let $s_1 = 72\epsilon^{-2}$ and $s_2 = 2\log \delta$.
We first computes $s_2$ random variables $Y_1, Y_2, \dots, Y_{s_2}$ and
outputs their median $\mathrm{Med}$, where each $Y_i$ is the average of $s_1$ random variables $Y_{ij}$: $1 \leq j \leq s_1$ and $Y_{ij}$ are independent, identically
distributed random variables computed by Figure~\ref{fig:algx}. By Chebyshev's inequality, we know that for any fixed $i$,
\begin{displaymath}
\Pr \big (\big | Y_i - \| \vec v \| \big | \big ) \geq \epsilon \|\vec v\|] \leq \frac{\mathrm{Var}(Y_i)}{\epsilon^2 \|\vec v\|^2} = \frac{(1/s_1)\mathrm{Var}[Y]}{\epsilon^2\|\vec v\|^2}
= \frac{(9\epsilon^2/72) \|\vec v\|^2}{\epsilon^2 \|\vec v\|^2} = \frac 1 8.
\end{displaymath}
Finally, by standard Chernoff bound arguments (see for example Chapter 4 of \cite{MU05}), the probability that more than $s_2/2$
of the variables $Y_i$ deviate by more than $\epsilon\|\vec v\|$ from $\|\vec v\|$ is at most $\delta$. In case this does not happen,
the median $\mathrm{Med}$ supplies a good estimate to the required quantity $\|\vec v\|$ as needed.
\end{proof}

\section{The General Case $k\geq 3$}\label{sec:estimator}

Now let us move to the general case where $k \geq 3$. Recall that $\vecv$ is a vector in $\mathrm R^{n^k}$ that maintains certain statistics of a data stream, and we are interested in estimating its $\ell_2$ norm $\|\vecv\|$. There is a natural generalization for Indyk and McGregor's method for $k = 2$ to construct an estimator for $\|\vecv\|$: let $h_1,\dots,h_k:[n] \rightarrow \{-1, 1\}$ be independent copies of 4-wise independent hash functions (namely, $h_i(1),\dots,h_i(n) \in \{-1,1\}$ are $4$-wise independent hash functions for each $i\in[k]$, and $h_1(\cdot),\dots,h_k(\cdot)$ are mutually independent.).  Let $H(\bfp) = \prod_{i =1}^k h_j(p_j)$. The estimator $Y$ is defined as $Y \equiv \left(\sum_{\bfp \in [n]^k}\vec v_{\bfp}H(\bfp)\right)^2$.

Our goal is to show that $\mathrm E[Y] = \|\vec v\|^2$ and $\mathrm{Var}[Y]$ is reasonably small so that a streaming algorithm maintaining multiple independent instances of estimator $Y$ will be able to output an approximately correct estimation of $\|\vecv\|$ with high probability. Notice that when $\|\vecv\|$ represents the $\ell_2$ distance between the joint distribution and the tensors of the marginal distributions, the estimator can be computed efficiently in a streaming model similarly to as in  Figure~\ref{fig:algx}. We stress that our result is applicable to a broader class of $\ell_2$-norm estimation problems, as long as the vector $\vecv$ to be estimated has a corresponding efficiently computable estimator $Y$ in an appropriate streaming model. Formally, we shall prove the following main lemma in the next subsection.


\begin{lemma} \label{lem:main_lemma} Let $\vecv $ be a vector in $\mathrm R^{n^k}$, and $h_1,\dots,h_k:[n] \rightarrow \{-1, 1\}$ be independent copies of 4-wise independent hash functions. Define $H(\bfp) = \prod_{i =1}^k h_j(p_j)$, and $Y \equiv \left(\sum_{\bfp \in [n]^k}\vec v_{\bfp}H(\bfp)\right)^2$. We have $\E[Y] = ||\vecv||$ and $\Var[Y] \leq 3^k \E[Y]^2$.
\end{lemma}

We remark that the bound on the variance in the above lemma is tight. One can verify that when the vector $\vecv$ is a uniform vector (i.e., all entries of $\vecv$ are the same), the variance of $Y$ is $\Omega(3^k E[Y]^2)$. With the above lemma, the following main theorem mentioned in the introduction immediately follows by a standard argument presented in the proof of Theorem \ref{thm:case2} in the previous section.

\begin{theorem}\label{tm:main1}
Let $\vec v$ be a vector in $\mathrm R^{[n]^k}$ that maintains an arbitrary statistics in a data stream of size $m$, in which every item is from $[n]^k$. Let $\epsilon,\delta \in (0,1)$ be real numbers. If there exists an algorithm that maintains an instance of $Y$ using $O(\mu(n,m,k, \epsilon, \delta))$ memory bits, then
there exists an algorithm $\Lambda$ such that:
\begin{enumerate}
\item With probability $\geq 1 - \delta$ the algorithm $\Lambda$ outputs a value between $[(1-\epsilon)\|\vecv\|^2, (1+\epsilon)\|\vecv|^2]$ and
\item the space complexity of $\Lambda$ is $O(3^k{1\over \epsilon^2}\log{1\over \delta}\mu(n,m,k, \epsilon, \delta))$.
\end{enumerate}
\end{theorem}

As discussed above, an immediate corollary is the existence of a one-pass space efficient streaming algorithm to detect the dependency of $k$ random variables in $\ell_2$-norm:

\begin{corollary}For every $\epsilon > 0$ and $\delta > 0$, there exists a randomized algorithm that computes,
given a sequence $a_1, \dots, a_m$ of $k$-tuples, in one pass and using
$O(3^k\epsilon^{-2}\log \frac 1 \delta(\log m + \log n))$ memory bits,
a number $Y$ so that the probability $Y$ deviates from the square of the $\ell_2$
distance between product and joint distribution by more than a factor of
$(1+\epsilon)$ is at most $\delta$.
\end{corollary}

\subsection{Analysis of the Sketch $Y$}\label{sec:analysis}

This section is devoted to prove Lemma \ref{lem:main_lemma}, where the main challenge is to bound the variance of $Y$.  The geometric approach of Indyk and McGregor \cite{IM08} presented in Section \ref{sec:kis2} for the case of $k=2$ can be extended to analyze the general case. However, we remark that the generalization requires new ideas. In particular, instead of performing ``local analysis'' that maps each rectangle to its diagonals, a more complex ``global analysis'' is needed in higher dimensions to achieve the desired bounds. The alternative proof we present here utilizes similar ideas, but relies on a more combinatorial rather than geometric approach.

For the expectation of $Y$, we have
\begin{eqnarray*}
\mathrm E[Y] & = & \mathrm E\left[\sum_{\bfp, \bfq \in [n]^k}\vecv_{\bfp}\cdot \vecv_{\bfq}\cdot H(\bfp) \cdot H(\bfq)\right] \\
& = & \sum_{\bfp \in [n]^k}\vec v_{\bfp}^2 \cdot \mathrm E\left[H(\bfp)^2\right] + \sum_{\bfp \neq \bfq \in [n]^k}\vecv_{\bfp} \cdot \vecv_{\bfq} \cdot \mathrm E\left[H(\bfp)H(\bfq)\right] \\
& = & \sum_{\bfp \in [n]^k}\vec v_{\bfp}^2 = ||\vecv||^2,
\end{eqnarray*}
where the last equality follows by $H(\bfp)^2 = 1$, and $\mathrm E\left[H(\bfp)H(\bfq)\right] =0$ for $\bfp \neq \bfq$.

Now, let us start to prove $\Var[Y] \leq 3^k \mathrm E[Y]^2$. By definition, $\Var[Y] = \mathrm E[ (Y- \E[Y])^2]$, so we need to understand the following random variable:
\begin{equation}\label{eq:X defined}
Err \equiv Y - \E[Y] = \sum_{\bfp \neq \bfq\in [n]^k} H(\bfp)H(\bfq)\vecv_{\bfp}\vecv_{\bfq}.
\end{equation}
The random variable $Err$ is a sum of terms indexed by pairs $(\bfp,\bfq) \in [n]^k \times [n]^k$ with $\bfp \neq \bfq$. At a very high level, our analysis consists of two steps. In the first step, we group the terms in $Err$ properly and simplify the summation in each group. In the second step,  we expand the square of the sum in $\Var[Y] = \mathrm E[Err^2]$ according to the groups and apply Cauchy-Schwartz inequality three times to bound the variance.



We shall now gradually introduce the necessary notation for grouping the terms in $Err$ and simplifying the summation. We remind the reader that vectors over the reals (e.g., $\vecv \in R^{n^k}$) are denoted by $\vecv, \vec{w},\vec{r}$, and vectors over $[n]$ are denoted by $\bfp,\bfq,\bfa,\bfb,\bfc,\bfd$ and referred as \emph{index vectors}.  We use $S\subseteq [k]$ to denote a subset of $[k]$, and let $\barS = [k] \backslash S$. We use $\mathrm{Ham}(\bfp, \bfq)$ to denote the \emph{Hamming distance} of index vectors $\bfp,\bfq \in [n]^k$, i.e., the number of coordinates where $\bfp$ and $\bfq$ are different. 


\begin{definition}\label{def:projection} \emph{(Projection and inverse projection)} Let $\bfc \in [n]^k$ be an index vector and $S \subseteq [k]$ a subset. We define the \emph{projection of $\bfc$ to $S$}, denoted by $\Phi_S(\bfc) \in [n]^{|S|}$, to be the vector $\bfc$ restricted to the coordinates in $S$.
Also, let $\bfa \in [n]^{|S|}$ and $\bfb \in [n]^{k - |S|}$ be index vectors. We define the \emph{inverse projection of $\bfa$ and $\bfb$ with respect to $S$}, denoted by $\Phi^{-1}_{S}(\bfa, \bfb) \in [n]^k$, as the index vector $\bfc \in [n]^k$ such that $\Phi_{S}(\bfc) = \bfa$ and $\Phi_{\barS}(\bfc) = \bfb$.
\end{definition}


We next define \emph{pair groups} and use the definition to group the terms in $Err$.

\begin{definition}\label{def:pairs projection} \emph{(Pair Group)}
Let $S \subseteq [k]$ be a subset of size $|S| = t$. Let $\bfc, \bfd\in [n]^t$ be a pair of index vectors with $\mathrm{Ham}(\bfc, \bfd) = t$ (i.e., all coordinates of $\bfc$ and $\bfd$ are distinct.). The \emph{pair group} $\sigma_S(\bfc, \bfd)$ is the set of pairs $(\bfp,\bfq) \in [n]^k \times [n]^k$ such that (i) on coordinate $S$, $\Phi_S(\bfp) = \bfc$ and $\Phi_S(\bfq) = \bfd$, and (ii) on coordinate $\barS$, $\bfp$ and $\bfq$ are the same, i.e., $\Phi_{\barS}(\bfp) = \Phi_{\barS}(\bfq)$. Namely,
\begin{equation}
\sigma_S(\bfc, \bfd) = \left\{(\bfp, \bfq) \in [n]^k \times [n]^k : \Big(\bfc = \Phi_S(\bfp)\Big) \wedge \Big(\bfd = \Phi_S(\bfq) \Big) \wedge \Big(\Phi_{\barS}(\bfp) = \Phi_{\barS}(\bfq)\Big)\right\}.
\end{equation}
\end{definition}

To give some intuition for the above definitions, we note that for every $\bfa \in [n]^{|\barS|}$, there is a unique pair $(\bfp,\bfq) \in \sigma_S(\bfc,\bfd)$ with $\bfa = \Phi_{\barS}(\bfp) = \Phi_{\barS}(\bfq)$, and so $|\sigma_S(\bfc,\bfd)| = n^{|\barS|}$. On the other hand, for every pair $(\bfp,\bfq) \in [n]^k \times [n]^k$ with $\bfp \neq \bfq$, there is a unique non-emtpy $S \subseteq [k]$ such that $\bfp$ and $\bfq$ are distinct on exactly coordinates in $S$. Therefore, $(\bfp,\bfq)$ belongs to exactly one pair group $\sigma_S(\bfc,\bfd)$. It follows that we can partition the summation in $Err$ according to the pair groups:
\begin{equation}\label{eq:errdecompose}
Err = \sum_{\substack{S \subseteq [k] \\S \neq \emptyset}}\ \ \ \ \  \sum_{\substack{\bfc,\bfd \in [n]^{|S|},\\ \mathrm{Ham}(\bfc, \bfd) = |S|}} \ \ \ \ \  \sum_{\substack{(\bfp,\bfq) \in \\ \sigma_S(\bfc,\bfd)}} H(\bfp)H(\bfq)\vecv_{\bfp}\vecv_{\bfq}.
\end{equation}

We next observe that for any pair $(\bfp,\bfq) \in \sigma_S(\bfc,\bfd)$, since $\bfp$ and $\bfq$ agree on coordinates in $\barS$, the value of the product $H(\bfp)H(\bfq)$ depends only on $S$, $\bfc$ and  $\bfd$. More precisely,
$$H(\bfp)H(\bfq) = \prod_{i\in[k]} h_i(p_i) h_i(q_i) = \left( \prod_{i\in S} h_i(p_i) h_i(q_i) \right) \cdot \left( \prod_{i\in \barS} h_i(p_i)^2 \right) = \prod_{i\in S} h_i(p_i) h_i(q_i),$$
which depends only on $S$, $\bfc$ and  $\bfd$ since  $\Phi_S(\bfp) = \bfc$ and $\Phi_S(\bfq) = \bfd$. This motivates the definition of \emph{projected hashing}.

\begin{definition}\label{def:projectedhashing} \emph{(Projected hashing)}
Let $S = \{s_1, s_2, \dots, s_t\}$ be a subset of $[k]$, where $s_1 < s_2 < \dots < s_j$.
Let $\bfc\in [n]^t$. We define the \emph{projected hashing} $H_S(\bfc) = \prod_{i \leq t}h_{s_i}(c_i)$.
\end{definition}
We can now translate the random variable $Err$ as follows:
\begin{equation}\label{eq: outside ht eprojected pairs}
Err = \sum_{\substack{S \subseteq [k] \\S \neq \emptyset}}\ \ \ \ \  \sum_{\substack{\bfc,\bfd \in [n]^{|S|},\\ \mathrm{Ham}(\bfc, \bfd) = |S|}} \left( H_S(\bfc)H_S(\bfd)  \sum_{\substack{(\bfp,\bfq) \in \\ \sigma_S(\bfc,\bfd)}} \vecv_{\bfp}\vecv_{\bfq} \right).
\end{equation}

Fix a pair group $\sigma_S(\bfc,\bfd)$, we next consider the sum $\sum_{\substack{(\bfp,\bfq) \in \sigma_S(\bfc,\bfd)}} \vecv_{\bfp}\vecv_{\bfq}$. Recall that for every $\bfa \in [n]^{|\barS|}$, there is a unique pair $(\bfp,\bfq) \in \sigma_S(\bfc,\bfd)$ with $\bfa = \Phi_{\barS}(\bfp) = \Phi_{\barS}(\bfq)$. The sum can be viewed as the inner product of two vectors of dimension $n^{|\barS|}$ with entries indexed by $\bfa \in [n]^{|\barS|}$. To formalize this observation, we introduce the definition of \emph{hyper-projection} as follows.

\begin{definition}\label{def:hyperproj} \emph{(Hyper-projection)} Let $\vecv \in R^{n^k}$, $S \subseteq [k]$, and $\mathbf \bfc \in [n]^{|S|}$. The \emph{hyper-projection} $\Upsilon_{S,\bfc}(\vecv)$ of $\vecv$ (with respect to $S$ and $\bfc$) is a vector $\vec{w} = \Upsilon_{S,\bfc}(\vecv)$ in $\mathrm R^{[n]^{k - |S|}}$ such that $\vec{w}_{\bfd} = \vecv_{\Phi^{-1}_S(\bfc, \bfd)}$ for all $\bfd \in [n]^{k - |S|}$.
\end{definition}

Using the above definition, we continue to rewrite the $Err$ as
\begin{equation}\label{eq: turn_to_inner_prod}
Err = \sum_{\substack{S \subseteq [k] \\S \neq \emptyset}}\ \ \ \ \  \sum_{\substack{\bfc,\bfd \in [n]^{|S|},\\ \mathrm{Ham}(\bfc, \bfd) = |S|}}  H_S(\bfc)H_S(\bfd) \cdot \langle \Upsilon_{S,\bfc}(\vecv),   \Upsilon_{S,\bfd}(\vecv) \rangle .
\end{equation}

Finally, we consider the product $H_S(\bfc)H_S(\bfd)$ again and introduce the following definition to further simplify  the $Err$.


\begin{definition}\label{def:similar pairs} \emph{(Similarity and dominance)}
Let $t$ be a positive integer.
\begin{itemize}
\item Two pairs of index vectors $(\bfc, \bfd) \in [n]^t \times [n]^t$ and $(\bfa, \bfb) \in [n]^t \times [n]^t$ are \emph{similar} if for all $i\in [t]$, the two sets $\{c_i, d_i\}$ and $\{a_i, b_i\}$ are equal. We denote this as $(\bfa, \bfb) \sim (\bfc, \bfd)$.

\item Let $\bfc$ and $\bfd \in [n]^t$ be two index vectors. We say $\bfc$ \emph{is dominated by} $\bfd$ if $c_i < d_i$ for all $i \in [t]$. We denote this as $\bfc \prec \bfd$. Note that $\bfc \prec \bfd \Rightarrow \mathrm{Ham}(\bfc, \bfd) = t$.
\end{itemize}
\end{definition}

Now, note that if $(\bfa, \bfb) \sim (\bfc, \bfd)$, then $H_S(\bfa)H_S(\bfb) = H_S(\bfc)H_S(\bfd)$ since the value of the product  $H_S(\bfc)H_S(\bfd)$ depends on the values $\{c_i, d_i\}$ only as a set. It is also not hard to see that  $\sim$ is an equivalence relation, and for every equivalent class $[(\bfa,\bfb)]$, there is a unique $(\bfc,\bfd) \in [(\bfa,\bfb)]$ with $\bfc \prec \bfd$. Therefore, we can further rewrite the $Err$ as
\begin{equation}\label{eq:final_err}
Err = \sum_{\substack{S \subseteq [k] \\S \neq \emptyset}}\  \sum_{\bfc \prec \bfd \in [n]^{|S|}}   H_S(\bfc)H_S(\bfd)\cdot \left( \sum_{(\bfa,\bfb) \sim (\bfc, \bfd)} \langle \Upsilon_{S,\bfa}(\vecv),   \Upsilon_{S,\bfb}(\vecv) \rangle \right).
\end{equation}

We are ready to bound the term $\mathrm E[Err^2]$ by expanding the square of the sum according to Equation (\ref{eq:final_err}). We first show in Lemma \ref{lem:vanish} below that all the cross terms in the following expansion vanish.

\begin{equation}\label{eq:expand_var}
\Var[Y] = \sum_{\substack{S,S' \subseteq [k] \\S, S' \neq \emptyset}}\  \sum_{\substack{\bfc \prec \bfd \in [n]^{|S|} \\  \bfc' \prec \bfd' \in [n]^{|S|'}}} \E[ H_S(\bfc)H_S(\bfd)H_{S'}(\bfc')H_{S'}(\bfd')]  \cdot
$$
$$
\left[\left(\sum_{(\bfa,\bfb) \sim (\bfc, \bfd)} \langle \Upsilon_{S,\bfa}(\vecv),   \Upsilon_{S,\bfb}(\vecv) \rangle \right)
\left(\sum_{(\bfa',\bfb') \sim (\bfc', \bfd')} \langle \Upsilon_{S',\bfa'}(\vecv),   \Upsilon_{S',\bfb'}(\vecv) \rangle \right)
\right].
\end{equation}

\begin{lemma} \label{lem:vanish} Let $S$ and $S'$ be subsets of $[k]$, and $\bfc \prec \bfd \in [n]^{|S|}$ and $\bfc' \prec \bfd' \in [n]^{|S'|}$ index vectors. We have $E[ H_S(\bfc)H_S(\bfd)H_{S'}(\bfc')H_{S'}(\bfd')] \in \{0,1\}$.
Furthermore, we have

\noindent
$\E[ H_S(\bfc)H_S(\bfd)H_{S'}(\bfc')H_{S'}(\bfd')]  = 1$ iff $(S = S') \wedge (\bfc = \bfc') \wedge (\bfd = \bfd')$.
\end{lemma}
\begin{proof} Recall that $h_1,\dots,h_k$ are independent copies of $4$-wise independent uniform random variables over $\{ -1, 1\}$. Namely, for every $i\in[k]$, $h_i(1),\dots,h_i(n)$ are $4$-wise independent, and $h_1(\cdot),\dots,h_k(\cdot)$ are mutually independent. Observe that for every $i\in[k]$, there are at most $4$ terms out of $h_i(1),\dots,h_i(n)$ appearing in the product $H_S(\bfc)H_S(\bfd)H_{S'}(\bfc')H_{S'}(\bfd')$. It follows that all distinct terms appearing in $H_S(\bfc)H_S(\bfd)H_{S'}(\bfc')H_{S'}(\bfd')$ are mutually independent uniform random variable over $\{-1, 1\}$. Therefore, the expectation is either 0, if there is some $h_i(j)$ that appears an odd number of times, or 1, if all $h_i(j)$ appear an even number of times. By inspection, the latter case happens if and only if  $(S = S') \wedge (\bfc = \bfc') \wedge (\bfd = \bfd')$.
\end{proof}

By the above lemma, Equation (\ref{eq:expand_var}) is simplified to
\begin{equation}\label{eq:simplified_var}
\Var[Y] = \sum_{\substack{S \subseteq [k]\\ S \neq \emptyset}}\  \sum_{\bfc \prec \bfd \in [n]^{|S|}}  \left( \sum_{(\bfa,\bfb) \sim (\bfc, \bfd)} \langle \Upsilon_{S,\bfa}(\vecv),   \Upsilon_{S,\bfb}(\vecv) \rangle\right)^2.
\end{equation}


We next apply the Cauchy-Schwartz inequality three times to bound the above formula. Consider a subset $S \subseteq [k]$ and a pair $\bfc \prec \bfd \in [n]^{|S|}$. Note that there are precisely $2^{|S|}$ pairs $(\bfa,\bfb)$ such that $(\bfa, \bfb)\sim (\bfc, \bfd)$. Thus, by the Cauchy-Schwartz inequality:
\begin{eqnarray*}
 \left(\sum_{\substack{(\bfa, \bfb) \in [n]^{|S|} \\(\bfa, \bfb)\sim (\bfc, \bfd)}} \langle\Upsilon_{S,\bfa}(\vecv), \Upsilon_{S,\bfb}(\vec v)\rangle\right)^2 &\leq& 2^{|S|} \sum_{\substack{(\bfa, \bfb) \in [n]^{|S|}\\ (\bfa, \bfb)\sim (\bfc, \bfd)}} (\langle\Upsilon_{S,\bfa}, \Upsilon_{S,\bfb}\rangle)^2 \\
 &\leq & 2^{|S|}\sum_{\substack{(\bfa,\bfb) \in [n]^{|S|}\\ (\bfa, \bfb)\sim (\bfc,\bfd)}} \langle\Upsilon_{S,\bfa}(\vecv), \Upsilon_{S,\bfa}(\vecv)\rangle\cdot\langle\Upsilon_{S,\bfb}, \Upsilon_{S, \bfb}(\vecv)\rangle.
\end{eqnarray*}
Notice that in the second inequality, we applied Cauchy-Schwartz in a component-wise manner.
Next, for a subset $S \subseteq [k]$, we can apply the Cauchy-Schwartz inequality a third time (from the third line to the fourth line) as follows:
\begin{eqnarray*}
& & \sum_{\bfc \prec \bfd \in [n]^{|S|}} \ \ \ \ \  \left(\sum_{\substack{(\bfa, \bfb) \in [n]^{|S|}\\ (\bfa, \bfb)\sim (\bfc, \bfd)}} \langle\Upsilon_{S,\bfa}(\vec v), \Upsilon_{S,\bfb}(\vec v)\rangle\right)^2 \\
& \leq & 2^{|S|}\sum_{\bfc \prec \bfd \in [n]^{|S|}} \ \ \ \sum_{\substack{(\bfa,\bfb) \in [n]^{|S|}\\ (\bfa, \bfb)\sim (\bfc,\bfd)}} \langle\Upsilon_{S,\bfa}(\vec v), \Upsilon_{S,\bfa}(\vec v)\rangle\cdot\langle\Upsilon_{S,\bfb}(\vec v), \Upsilon_{S, \bfb}(\vec v)\rangle \\
& = & 2^{|S|}\sum_{\substack{\bfc, \bfd \in [n]^{|S|}\\ \mathrm{Ham}(\bfc, \bfd) = |S|}} \langle\Upsilon_{S,\bfc}(\vec v), \Upsilon_{S,\bfc}(\vec v)\rangle\cdot\langle\Upsilon_{S, \bfd}(\vec v), \Upsilon_{S, \bfd}(\vec v)\rangle \\
& \leq & 2^{|S|}\sum_{\bfc,\bfd \in [n]^{|S|} } \langle\Upsilon_{S,\bfc}(\vec v), \Upsilon_{S,\bfc}(\vec v)\rangle\cdot\langle\Upsilon_{S, \bfd}(\vec v), \Upsilon_{S, \bfd}(\vec v)\rangle \\
& = & 2^{|S|}\left(\sum_{\bfc \in [n]^{|S|}} \langle\Upsilon_{S,\bfc}(\vec v), \Upsilon_{S,\bfc}(\vec v)\rangle\right)^2.
\end{eqnarray*}

Finally, we note that by definition,  we have $\sum_{\bfc \in [n]^{|S|}} \langle\Upsilon_{S,\bfc}(\vec v), \Upsilon_{S,\bfc}(\vec v)\rangle = ||\vecv||^2$, which equals to $\mathrm E[Y]$. It follows that the variance in Equation (\ref{eq:simplified_var}) can be bounded by
$$\Var[Y] \leq \sum_{S \subseteq [k], S \neq \emptyset} 2^{|S|} \cdot \mathrm E[Y]^2 = \mathrm E[Y]^2 \sum_{i=1}^k {k \choose i}2^i = (3^k-1)\mathrm E[Y]^2,$$
which finishes the proof of Lemma \ref{lem:main_lemma}.

\section{Conclusion}

There remain several open questions left in this space.  Lower bounds,
particularly bounds that depend non-trivially on the dimension $k$,
would be useful.  There may still be room for better algorithms for
testing $k$-wise independence in this manner using the $\ell_2$ norm.
A natural generalization would be to find a
particularly efficient algorithm for testing $k$-out-of-$s$-wise
independence (other than handling each set of $k$ variable
separately).  More generally, a question given in \cite{IM08}, to
identify random variables whose correlation exceeds some threshold
according to some measure, remains widely open.

\bibliographystyle{plain}
{\footnotesize
\bibliography{braverman}
}
%
%
\vspace{-1cm}

\end{document}